\documentclass[prl,superscriptaddress,amsmath,twocolumn,amssymb]{revtex4}

\usepackage{xr}

\usepackage{subfigure}
\usepackage{amsmath}
\usepackage{pgfplots}
\pgfplotsset{compat=1.5}
\usepgfplotslibrary{groupplots}

\usepackage{makecell}

\usepackage[normalem]{ulem}

\usepackage{multirow}
\usepackage{placeins}
\usepackage{subfigure}
\usepackage{soul}
\usepackage{color,xcolor}
\usepackage[colorlinks,linkcolor=red,anchorcolor=blue,citecolor=blue,urlcolor=blue]{hyperref}
\usepackage{cleveref}

\newcommand{\svd}{{\rm SVD}}

\newcommand{\spacecpx}{\mathcal{C}^s}
\newcommand{\timecpx}{\mathcal{C}^t}
\newcommand{\bA}{{\mathbf{A}}}
\newcommand{\bE}{{\mathbf{E}}}

\newcommand{\istd}{Information Systems Technology and Design, Singapore University of Technology and Design, 8 Somapah Road, 487372 Singapore}
\newcommand{\sutd}{Science and Math Cluster and EPD Pillar, Singapore University of Technology and Design, 8 Somapah Road, 487372 Singapore}

\newcommand{\zz}{Henan Key Laboratory of Quantum Information and Cryptography, SSF IEU, Zhengzhou 450001, China}
\newcommand{\sft}{Quantum Intelligence Lab (QI-Lab), Supremacy Future Technologies (SFT), Guangzhou 511340, China}
\newcommand{\pc}{Center for Quantum Computing, Peng Cheng Laboratory, Shenzhen 518055, China}

\begin{document}

\title{General-purpose quantum circuit simulator with Projected Entangled-Pair States and the quantum supremacy frontier}

\author{Chu Guo}
\thanks{These authors contribute equally to this work.}
\affiliation{\zz}

\author{Yong Liu}
\thanks{These authors contribute equally to this work.}
\affiliation{Institute for Quantum Information \& State Key Laboratory of High Performance Computing, College of Computer, National University of Defense Technology, Changsha 410073, China}
\author{Min Xiong}
\affiliation{Institute for Quantum Information \& State Key Laboratory of High Performance Computing, College of Computer, National University of Defense Technology, Changsha 410073, China}
\author{Shichuan Xue}
\affiliation{Institute for Quantum Information \& State Key Laboratory of High Performance Computing, College of Computer, National University of Defense Technology, Changsha 410073, China}

\author{Xiang Fu}
\affiliation{Institute for Quantum Information \& State Key Laboratory of High Performance Computing, College of Computer, National University of Defense Technology, Changsha 410073, China}
\author{Anqi Huang}
\affiliation{Institute for Quantum Information \& State Key Laboratory of High Performance Computing, College of Computer, National University of Defense Technology, Changsha 410073, China}
\author{Xiaogang Qiang}
\affiliation{Institute for Quantum Information \& State Key Laboratory of High Performance Computing, College of Computer, National University of Defense Technology, Changsha 410073, China}
\author{Ping Xu}
\affiliation{Institute for Quantum Information \& State Key Laboratory of High Performance Computing, College of Computer, National University of Defense Technology, Changsha 410073, China}

\author{Junhua Liu}
\affiliation{\istd}
\affiliation{\sft}

\author{Shenggen Zheng}
\affiliation{\pc}

\author{He-Liang Huang}
\affiliation{\zz}
\affiliation{Hefei National Laboratory for Physical Sciences at Microscale and Department of Modern Physics,\\
University of Science and Technology of China, Hefei, Anhui 230026, China}
\affiliation{CAS Centre for Excellence and Synergetic Innovation Centre in Quantum Information and Quantum Physics,\\
University of Science and Technology of China, Hefei, Anhui 230026, China}

\author{Mingtang Deng}
\affiliation{Institute for Quantum Information \& State Key Laboratory of High Performance Computing, College of Computer, National University of Defense Technology, Changsha 410073, China}

\author{Dario Poletti}
\email{dario\_poletti@sutd.edu.sg}
\affiliation{\sutd}
\author{Wan-Su Bao}
\email{bws@qiclab.cn}
\affiliation{\zz}
\affiliation{CAS Centre for Excellence and Synergetic Innovation Centre in Quantum Information and Quantum Physics,\\
University of Science and Technology of China, Hefei, Anhui 230026, China}

\author{Junjie Wu}
\email{junjiewu@nudt.edu.cn}
\affiliation{Institute for Quantum Information \& State Key Laboratory of High Performance Computing, College of Computer, National University of Defense Technology, Changsha 410073, China}

\begin{abstract}
Recent advances on quantum computing hardware have pushed quantum computing to the verge of quantum supremacy. Here we bring together many-body quantum physics and quantum computing by using a method for strongly interacting two-dimensional systems, the Projected Entangled-Pair States, to realize an effective general-purpose simulator of quantum algorithms. The classical computing complexity of this simulator is directly related to the entanglement generation of the underlying quantum circuit rather than the number of qubits or gate operations. We apply our method to study random quantum circuits, which allows to quantify precisely the memory usage and the time requirements of random quantum circuits. We demonstrate our method by computing one amplitude for a $7\times 7$ lattice of qubits with depth $(1+40+1)$ on the Tianhe-2 supercomputer.
\end{abstract}

\date{\today}

\maketitle

Quantum computers offer the promise of efficiently solving certain problems that are intractable for classical computers, most famously factorizing large numbers~\cite{Feynman1982, Shor1994, BoixoNeven2017}. With the rapid progress of various quantum systems towards Noisy Intermediate-Scale Quantum computing devices~\cite{LundRalph2017,Huang2017,zhang2017,Huang2018, Wright2019,Kelly2019,Gong2019,Wang2019}, we are now on the verge of \emph{quantum supremacy}~\cite{Preskill2012}, i.e. demonstrating that a quantum computer has the ability to do a computation that no classical computers can tackle, an important milestone in the field of computer science. Various candidates have been suggested to demonstrate quantum supremacy, such as BosonSampling~\cite{Aaronson2011,Wu2018}, the instantaneous quantum polynomial protocol~\cite{Shepherd2009,Bremner2011} and random quantum circuits (RQCs)~\cite{BoixoNeven2017,BoulandVazirani2018} which demand less physical resources and are easier to implement compared to, for instance, factorization.

A central aspect for all these near-term supremacy proof-of-principle computations is to produce a quantum state using as fewer number of qubits as well as quantum gate operations as possible, which would nevertheless be highly entangled and hence difficult to obtain and/or characterize by a classical computer, for instance by sampling from it in the computational basis. In the meanwhile, it is important to find effective ways to simulate accurately quantum algorithms on classical computers, which could be used as a benchmarking baseline and to validate near term quantum devices. In the field of quantum many-body physics, tensor network states are often used to efficiently represent quantum states with a sizeable amount of entanglement~\cite{Eisert2013, Orus2014}. The storage required by these tensor network states is closely related to the amount of entanglement of the quantum state. Recently, matrix product states (which are one-dimensional tensor networks) have been applied to simulate quantum circuits~\cite{McCaskeyHumble2018}. However, the performance of matrix product states is much less effective if the underlying quantum system is essentially two-dimensional. In this work, we present an efficient and generic quantum circuit simulator based on the Projected Entangled-Pair States (PEPS)~\cite{VerstraeteCirac2004, VerstraeteCirac2006, MurgCirac2006, JordanCirac2008, GuWen2008, JiangXiang2008, XieXiang2009, MurgCirac2009, Orus2014}, a type of tensor-network quantum states representation designed for two-dimensional lattices.
Our PEPS-based simulator is a general-purpose quantum circuit simulator for arbitrary quantum circuits: it stores the full quantum state and it can be readily used to compute single amplitudes, observables, and also perform sequences of quantum measurements.

While the quantum circuit simulator we present can tackle generic circuits, in the following we focus on RQCs. They consist of a series of single and two-qubit gates which are applied to different qubits in a particular order. A group of commuting gates, which can be applied simultaneously, constitutes one layer of the circuit, and the more groups of operations that do not commute, the {\it deeper} the circuit is.
More precisely, for the depth of a circuit we will use the notation $(1+d+1)$ where the $ '1's $ indicate the Hadamard gates applied to each site at the beginning and at the end of the calculations, while $d$ is the number of non-commuting layers including controlled-Z (CZ) gates and single qubit gates applied to different sites.
RQCs are the standard benchmark for quantum supremacy as put forth by \cite{BoixoNeven2017}. The general complexity of quantum supremacy experiments is studied in~\cite{AaronsonChen2016}. For RQCs, it was previously shown in~\cite{BoixoNeven2018} that the complexity scales exponentially with $\min(O(dL_h),O(N))$.

RQCs have thus stimulated the search for efficient classical algorithms which would show where exactly the limits of classical simulations are~\cite{HanerSteiger2017, BoixoNeven2018, ChenGuo2018, BoulandVazirani2018, LiYang2018, PednaultWisnieff2018, ChenPan2019, MarkovBoixo2018, Villalonga2018, Villalonga2019}.
State of the art algorithms can be mainly divided into two categories. i) State-vector approach which stores the quantum state as a vector and evolves it directly. For example, in~\cite{HanerSteiger2017} a 45-qubit simulation is reported based on this approach. However, this approach is limited by the number of qubits due to the exponential growth of the Hilbert space. ii) Tensor-based approach, which represents the quantum states as tensors and specifies the input and output states as rank-1 Kronecker projectors. This approach is less sensitive to the number of qubits and has been pursued more actively. For instance, a full amplitude simulation of a $7\times 7$ circuit to depth $(1+39+1)$ was implemented in $4.2$ hours on Sunway TaihuLight supercomputer~\cite{LiYang2018}, which however exploits the weakness in the original design of RQCs in~\cite{BoixoNeven2017}. Recently, it was proposed to trade circuit fidelity for computational efficiency so as to match the fidelity of a given quantum computer~\cite{MarkovBoixo2018, Villalonga2018}, and practically compute around $1$ million amplitudes of a $7\times 7$ circuit to depth $(1+40+1)$ with $0.5\%$ circuit fidelity in $2.44$ hours on Summit supercomputer~\cite{Villalonga2019}. Our approach differs from the above approaches in that we use PEPS as the data structure to represent the quantum states. Quantum gate operations as well as quantum projections are adapted accordingly to this new data structure.

\begin{figure}[htbp]
  \centering
  \includegraphics[width=0.48\textwidth]{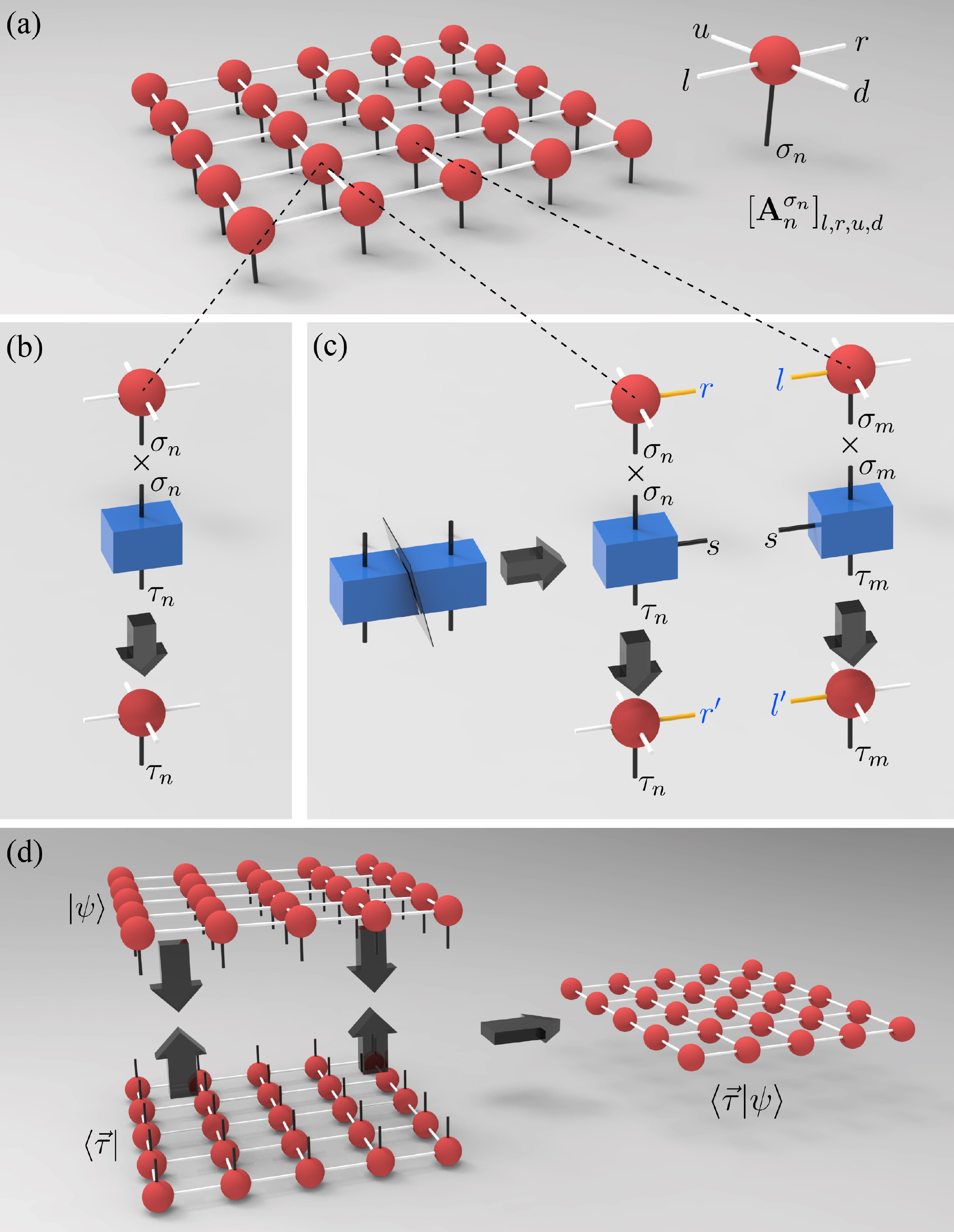}
  \caption{(a) PEPS on a $5\times 5$ lattice, each qubit of the lattice is represented with a rank-$5$ tensor $\left[\bA_n^{\sigma_n}\right]_{l,r,u,d}$, where $\sigma_n=0,1$ labels the physical dimension and $l,r,u,d$ label the auxiliary dimensions which connect $\left[\bA_n^{\sigma_n}\right]_{l,r,u,d}$ to the tensors on the neighbouring sites. (b) single-qubit gate operation on the PEPS. (c) Two-qubit gate operation on PEPS. (d) Overlapping of two PEPSs by contraction of all the physical dimensions of the two PEPSs and all the auxiliary dimensions inside each PEPS.}\label{fig:peps}
\end{figure}

{\bf {{Quantum Circuit Simulator Based on PEPS}}}. In the following we consider a two-dimensional rectangular lattice of size $L_v\times L_h$, where $L_v$ and $L_h$ are, respectively, the sizes in the vertical and horizontal directions. We use $N=L_vL_h$ to denote the total number of qubits. The quantum state on such a lattice can be represented as a PEPS~\cite{VerstraeteCirac2004, MurgCirac2006, JordanCirac2008}
\begin{align}\label{eq:PEPS}
\vert \psi \rangle = \sum_{\sigma_1, \dots, \sigma_N}
\mathcal{F}(\bA_1^{\sigma_1}\bA_2^{\sigma_2}\cdots\bA_N^{\sigma_N}) \vert \sigma_1, \sigma_2, \dots, \sigma_N \rangle,
\end{align}
where $\bA_n^{\sigma_n}$ is a rank-$5$ tensor with elements $\left[\bA_n^{\sigma_n}\right]_{l,r,u,d}$ at site $n$, with $\sigma=0,1$ corresponding to the physical dimension, and $l, r, u, d$ corresponding to the left, right, up and down auxiliary dimensions, see Fig.~\ref{fig:peps}(a). The function $\mathcal{F}$ in Eq.~(\ref{eq:PEPS}) indicates the sum over the common auxiliary indices. The bond dimension $\chi$ is defined as the maximum size of the four auxiliary dimensions,
\begin{align}
\chi = \max\{\dim(l), \dim(r), \dim(u), \dim(d)\},
\end{align}
and it characterizes the size of the PEPS.

In the language of PEPS, a single-qubit gate operation $U^{\tau_n}_{\sigma_n}$ on site $n$ only operates locally on the $n$-th tensor $\bA_n^{\sigma_n}$ (shown in Fig.~\ref{fig:peps}(b)), which can be written as
\begin{align}\label{eq:onebody}
\left[\bA_n^{\!\prime \;\tau_n}\right]_{l,r,u,d} = \sum_{\sigma_n} U^{\tau_n}_{\sigma_n}\left[\bA^{\sigma_n}_n\right]_{l,r,u,d}.
\end{align}
As we can see from Eq.~(\ref{eq:onebody}), the size of the local tensor is not affected by a single-qubit gate operation. For a two-qubit gate acting on a horizontally nearest-neighbour pair of qubits $(n, m)$ (shown in Fig.~\ref{fig:peps}(c)), denoted as $O_{\sigma_n, \sigma_m}^{\tau_n, \tau_m}$, we first use a by singular value decomposition (SVD) to factorize it into a product of two local tensors
\begin{align} \label{eq:factorize}
\svd(O_{\sigma_n, \sigma_m}^{\tau_n, \tau_m}) = \sum_s U_{\sigma_n, s}^{\tau_n} V_{s, \sigma_m}^{\tau_m},
\end{align}
where the singular values have been absorbed into $U$. 
The size of the auxiliary dimension $s$ is denoted as $\chi_o$, which, for any two-qubit controlled gate, is $\chi_o=2$.
The two local tensors $U$ and $V$ are then applied on the two qubits $n$ and $m$ separately, as single-qubit gate operations
\begin{align}
\left[\bA_n^{\!\prime\;\tau_n}\right]_{l,r',u,d} &= \sum_{\sigma_n} U^{\tau_n}_{\sigma_n, s}\left[\bA_n^{\sigma_n}\right]_{l,r,u,d}, \label{eq:twoqubita} \\
\left[\bA_m^{\!\prime\;\tau_m}\right]_{l',r,u,d} &= \sum_{\sigma_m} V^{\tau_m}_{s, \sigma_m}\left[\bA_m^{\sigma_m}\right]_{l,r,u,d}. \label{eq:twoqubitb}
\end{align}
Here we haved used the indices $r'=(r, s)$, $l'=(s, l)$, which bundles the two tensor dimensions into one. As a result, $\chi$ increases by a factor of $\chi_o$. To keep $\chi$ in a affordable size, one would usually use a subsequent singular value decomposition to compress the resulting tensors by throwing away singular values below a suitably chosen threshold. However, we point out that for RQCs we cannot perform such a compression because the distribution of the singular values after the two-qubit gate operation is almost flat, making it impossible for compression (this is also an indication that this problem has large entanglement across the whole circuit). Calculating a single amplitude of the final state $\vert\psi\rangle$ is done by projecting $\vert\psi\rangle$ onto a separable PEPS which encodes one spin configuration $\vert\vec{\tau}\rangle$, and then contracting the resulting tensor network, which can be written as
\begin{align} \label{eq:overlap}
\langle \vec{\tau} \vert \psi\rangle = \mathcal{F}(\bE_1\bE_2\cdots\bE_N),
\end{align}
where the rank-$4$ tensor $\left[\bE_n\right]_{l,r,u,d} = \left[\bA_n^{\sigma_n=\tau_n}\right]_{l,r,u,d}$. This calculations are depicted in Fig.~\ref{fig:peps}(d). To this end, we also note that with our method it is also straightforward to simulate sequences of quantum measurements. Concretely, to measure an $N$-th qubit system, we can first compute the probability that a qubit is in state $\vert0\rangle$ or $\vert1\rangle$. Then, we use another copy of the wavefunction (which is stored as PEPS), project the measured qubit in the relevant state, measure another qubit and so forth. In between different measurements more gates can be applied too, all seamlessly because we can effectively and efficiently compute and store the wavefunction of the system.

{\bf { {Application to random quantum circuits and complexity analysis}}}. In the following, we apply our PEPS simulator to study the two-dimensional RQCs of~\cite{GitHub,supplementary}.
The simulation of this circuit is divided into two parts: (i) circuit evolution and (ii) computing the overlap with randomly selected spin configurations, namely calculating the amplitudes. To quantify the size of the bond dimension required by the tensors, we realize that a single-qubit operation does not affect the size of the tensor it operates on, while a nearest-neighbour two-qubit controlled operation increases the sizes of the two tensors it operates on by a factor of $2$ as shown previously~\cite{iSWAP}.
This results in
\begin{align}
\chi \leq 2^{\lceil d/8\rceil}, \label{eq:chi}
\end{align}
where $\lceil\dots\rceil$ is the ceiling function.
The equality in Eq.~(\ref{eq:chi}) is reached if the depth $d$ can be divided by $8$ (each nearest-neighbour pair of sites will be acted on by a CZ gate in every $8$ depths). As can be seen from Eqs.~(\ref{eq:onebody}, \ref{eq:twoqubita}, \ref{eq:twoqubitb}), the cost of each gate operation on PEPS scales as $O(\chi^4)$, which is relatively cheap. As a result, circuit evolution can be performed very efficiently. In fact, we can simulate the exact evolution of a $12\times12$ lattice to a depth $(1+40+1)$ within minutes on a personal laptop.

In contrast, a well-known result about PEPS is that exactly computing the overlap as in Eq.~(\ref{eq:overlap}) is an exponentially hard problem~\cite{SchuchCirac2007}. While there exist approximate algorithms to evaluate Eq.~(\ref{eq:overlap}) which scale polynomially with $\chi$~\cite{VerstraeteCirac2006, JiangXiang2008, GuWen2008}, they are inadequate for RQCs due to the large entanglement of the states produced. In the following we ignore both the space and time complexity of circuit evolution and only focus on calculating one amplitude, since the cost of the former stage is negligible compared to the latter.

We have developed different strategies to evaluate Eq.~(\ref{eq:overlap}) efficiently, depending on the shape and size of the lattice. A generic strategy which works for any rectangular lattice has space and time complexities (assuming $L_v \geq L_h$) given by
\begin{align}
\spacecpx(L_v\times L_h\times d) &= 2^{\lceil d/8\rceil \left(L_h+1\right)}, \label{eq:space} \\
\timecpx(L_v\times L_h\times d) &= (L_h-2)(L_v-2)2^{\lceil d/8\rceil (L_h+3)}. \label{eq:time}
\end{align}
For square lattices, specialized tensor contraction strategies can be used to further reduce the complexity or for better parallelization (see~\cite{supplementary} for details of these strategies). We highlight here that Eqs.~(\ref{eq:space},\ref{eq:time}) are more accurate estimates for space and time complexities compared to the results of~\cite{BoixoNeven2018}, and the exact value will depend on the details of the particular implementation on the hardware. However, these numbers can work as a theoretical approximate benchmarking baseline for achieving quantum supremacy.

To give more precise numbers, using Eqs.~(\ref{eq:space},\ref{eq:time}) we can evaluate that simulating a $8\times 8$ lattice to a depth $(1+40+1)$ (same space complexity of a $10\times 10$ circuit to a depth $(1+32+1)$) would require $32$ TB of memory, while simulating a $8\times l$ (with $l > 8$) lattice to a depth $(1+40+1)$ would require about $0.5$ PB memory. However, simulating a $9\times 9$ lattice with a depth $(1+40+1)$ would require $16$ PB (petabytes) memory and simulating a $12\times 12$ lattice to a depth $(1+32+1)$ would require $8$ PB memory, which are currently out of reach. Our circuit simulator can straightforwardly be extended to other types of two-dimensional lattices including Google Bristlecone QPU architecture. By applying a complexity analysis to this architecture, we find that it only requires less than a manageable $0.6$ PB of memory to simulate an RQC with $72$ qubits at depth $(1+32+1)$ (for details of this analysis see~\cite{supplementary}).

\begin{table}[!htb]
\centering
\caption{{\bf Large-scale simulation with PEPS based circuit simulator.} The column denoted by ``Node usage'' indicates the number of cores used divided by the total available on Tianhe-2, and the corresponding percentage. ``Qubits'' and ``Depth'' describe the circuit analyzed while ``Elapsed time'' shows the time required to compute one amplitude.}
\label{TAB:BenchResult}
\begin{tabular}{p{2.8cm}p{1.5cm}lr}
\Xhline{1pt}
Node usage                          & Qubits        & Depth     &   Elapsed time\\
\hline
\multirow{3}*{4096/17920, 22\% }    & $7\times 7$   & (1+39+1)  &   9 min       \\
               ~                    & $7\times 7$   & (1+40+1)  &   31 min      \\
               ~                    & $8\times 8$   & (1+37+1)  &   68 min      \\
2048/17920, 11\%                    & $9\times 9$   & (1+31+1)  &   22 min      \\
1024/17920, 5\%                     & $10\times 10$ & (1+26+1)  &   9 min       \\
\Xhline{1pt}
\end{tabular}
\end{table}

\begin{figure}
\includegraphics[width=\columnwidth]{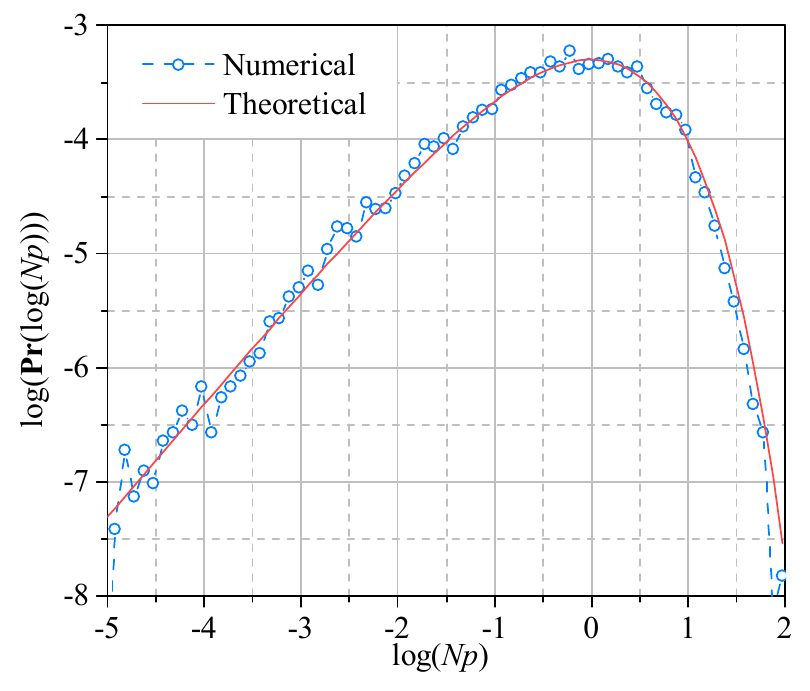}
\caption{ The blue circles show the log transformed probabilities from calculating $10000$ amplitudes, while the red line is the log transformed Porter-Thomas distribution. The circuit size is $8\times 8$ with a depth $(1+25+1)$.
} \label{fig:personal}
\end{figure}

To demonstrate the performance of our method, we have implemented small scale simulations on a personal computer, which takes less than $1$ hour to compute one amplitude of a $8\times 8$ circuit to a depth $(1+25+1)$ for a machine with $2$ cores of $2.8$ GHz frequency and $16$ GB memory. We computed $10000$ amplitudes then plotted the frequency with which each probability of configurations appear. This is represented in Fig.~\ref{fig:personal} by blue circles while the red continuous line shows the Porter-Thomas distribution, which is what is expected theoretically.

Our PEPS-based method can be readily scaled up onto a massive parallel computing platform. We implemented the large scale tensor contractions  based on an open-source software package Cyclops Tensor Framework~\cite{Solomonik2014}. The massive parallel benchmarking was executed on the Tianhe-2 supercomputer~\cite{Liao2014}. We have simulated a $7\times 7$ circuit with depth $(1+40+1)$ and a $10\times 10$ circuit with depth $(1+26+1)$. The simulation of the $7\times 7\times (1+40+1)$ circuit was done on 4096 nodes (22\%) of Tianhe-2, taking 31 minutes and 92.51 TB memory in total~\cite{supplementary}. Our large-scale simulation results are listed in TABLE~\ref{TAB:BenchResult}.

{\bf Conclusions}. In this work we have adapted the Projected Entangled-Pair States representation of quantum states from many-body quantum physics to build a general-purpose quantum circuit simulator. This simulator can be used to store effectively highly entangled wavefunctions, and it is readily adaptable to compute expectation values or simulate sequential quantum measurements. With this circuit simulator, we have computed an accurate estimate for the space and time complexity analysis of a standard random quantum circuit~\cite{GitHub}. Based on this analysis, we point out that simulating an $8\times l$ circuit to a depth $(1+40+1)$ or a Bristlecone-$72$ circuit to a depth $(1+32+1)$ are within reach of current supercomputing platforms.

We have implemented numerical experiments on a personal computer with a $8\times 8$ circuit to a depth $(1+25+1)$, and on Tianhe-2 supercomputer with a $10\times 10$ circuit to a depth $(1+26+1)$, as well as a $7\times 7$ circuit to a depth $(1+40+1)$. Currently we compute the amplitudes exactly, 
however we could also investigate the trade-off between fidelity and speed, so as to be able to sample many trajectories. For instance, we could reduce the memory requirement of our method by using the `cut' technique in~\cite{Villalonga2019}, namely mapping a large tensor contraction into summations over many smaller tensor contractions by unraveling several for-loops. More importantly, PEPS-based techniques which are currently used in quantum many-body physics can be transferred to the study of quantum circuits, for example for contractions and the evaluation of expectation values \cite{LubaschBanuls2014}. These investigations, which could be particularly useful for circuits in which the wavefunction can be effectively compressed, are left for future works, together with the plan to include the effects of noise or errors in order to characterize more closely the actual behavior of a noisy intermediate-scale quantum computer.

\begin{acknowledgments}
We gratefully acknowledge the help from China Greatwall Technology and National Supercomputing Center in Guangzhou. We thank Sergio Boixo and Giacomo Nannicini for helpful discussions. C. G. acknowledges support from National Natural Science Foundation of China under Grants No. 11504430 and No. 11805279. H.-L. H. acknowledges support from the Open Research Fund from State Key Laboratory of High Performance Computing of China (Grant No. 201901-01), National Natural Science Foundation of China under Grants No. 11905294, and China Postdoctoral Science Foundation. D.P. acknowledges support from the Singapore Ministry of Education, Singapore Academic Research Fund Tier-II (project MOE2016-T2-1-065). J.W. acknowledges support from National Natural Science Foundation of China under Grants No. 61632021.
\end{acknowledgments}


\clearpage
\onecolumngrid

\setcounter{section}{0}
\setcounter{equation}{0}
\setcounter{figure}{0}
\setcounter{table}{0}
\renewcommand{\theequation}{S\arabic{equation}}
\renewcommand{\thefigure}{S\arabic{figure}}
\renewcommand{\thetable}{S\Roman{table}}

\begin{center}\bf\large
    Supplementary Materials\\ General-purpose quantum circuit simulator with Projected Entangled-Pair States and the quantum supremacy frontier
\end{center}

\section{Introduction to Random Quantum Circuits}\label{app:googlecircuit}

\begin{figure}[h]
\includegraphics[width=1\columnwidth]{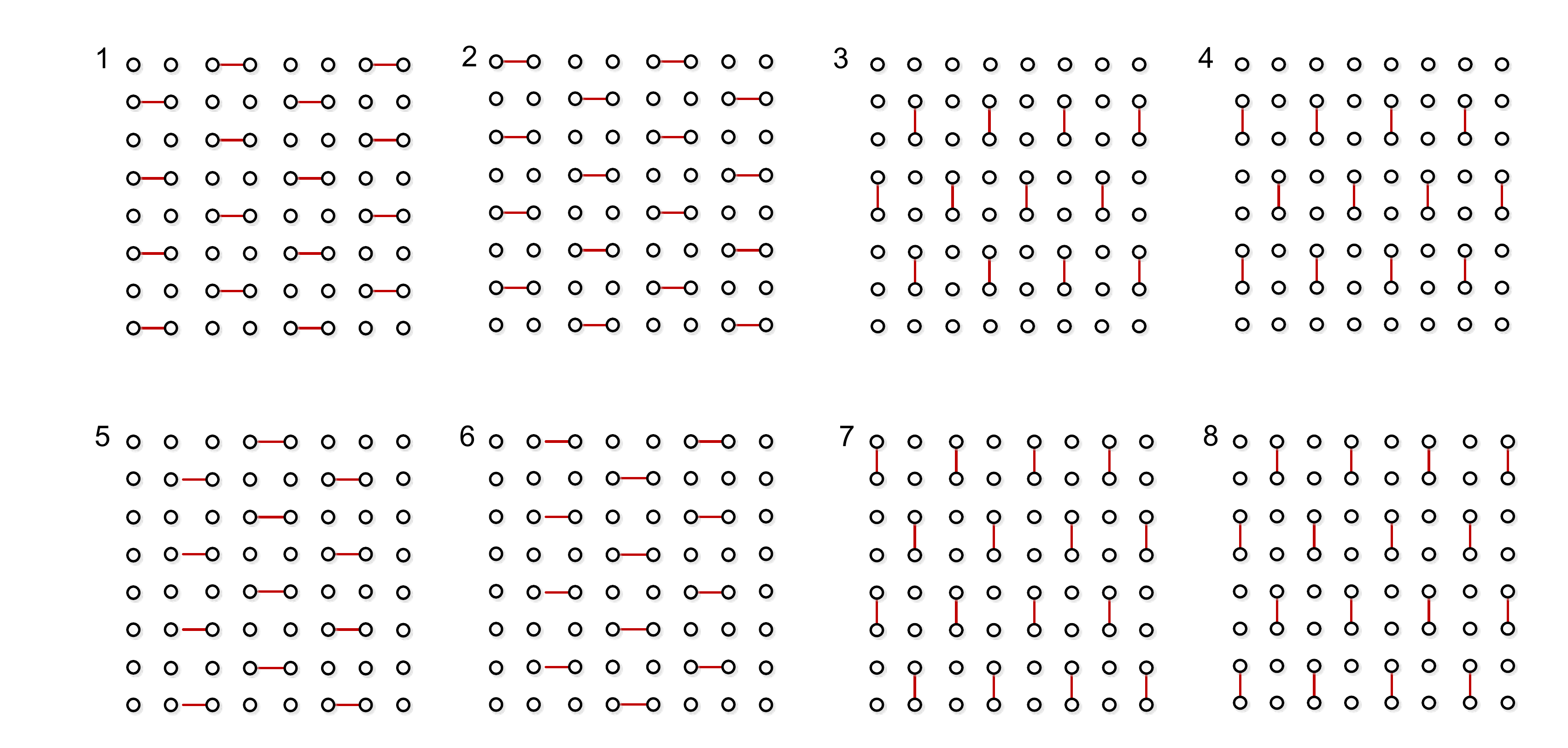}
\caption{ Layout of the CZ gates for the rank-2 random quantum circuit.
} \label{fig:twoDcircuit}
\end{figure}

For a $L_v \times L_h$ qubit lattice, the Random Quantum Circuit (RQC) defined by~\cite{BoixoNeven2017} is described as follows:
\begin{enumerate}
  \item Apply a Hadamard gate to each qubit to initialize the qubits to a symmetric superposition.
  \item Apply controlled-phase (CZ) gates alternating between eight configurations similar to Fig.~\ref{fig:twoDcircuit} to entangle neighbouring qubits.
  \item Apply a randomly chosen gate ($\text{T}$, $\text{X}^{1/2}$ or $\text{Y}^{1/2}$) to each qubit on which the CZ gates has not just been applied, according to the rules in~\cite{BoixoNeven2017}.
  \item Repeat steps 2 and 3 to add layers of depth to the circuit.
  \item Apply a final Hadamard gate to each qubit.
\end{enumerate}

It has been proven that this random quantum circuit satisfies both average-case hardness and anti-concentration condition~\cite{BoulandVazirani2018}, and hence it cannot be efficiently simulated on a classical computer.

\section{Algorithm for Exact Computation of the Overlap}\label{app:compute_overlap}

\begin{figure*}[!htb]
  \centering
  \includegraphics[width=0.95\textwidth]{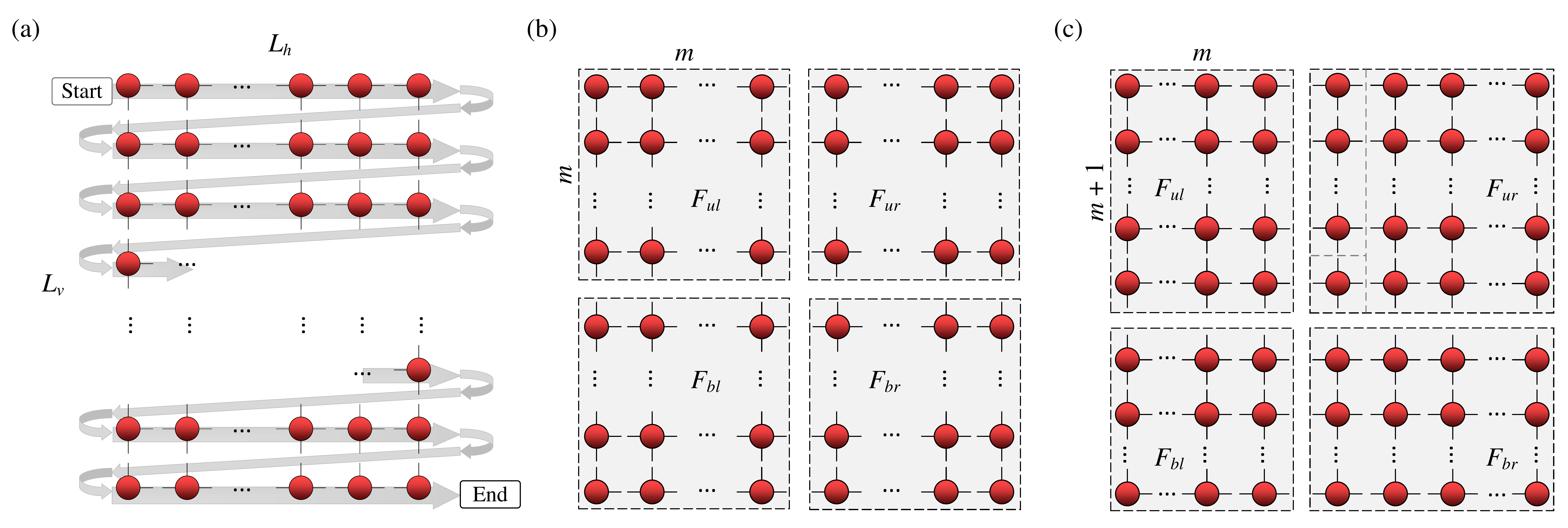}
  \caption{{\bf Contracting strategies for different lattices.} (a) Generic contracting scheme for lattices with $L_v \geq L_h$. One first contracts the tensors on each horizontal line from top down. For the case with $L_v < L_h$, one can contract the tensors on each vertical lines for left to right. In this strategy, the largest stored tensor is of rank $\min(L_h, L_v)+1$. (b) For a square lattice with an even number of qubits, namely $L_h=L_v=2m$, the tensor network is divided into four square sub-lattices with sizes $m\times m$. Each sub-lattice is contracted to get a single larger tensor, and then they are contracted to get the probability amplitude. In this strategy, the largest stored tensor has $\sqrt{N}$ indices. (c) For a square lattice with an odd number of qubits, namely $L_h=L_v=2m+1$, the tensor network is divided into four sub-lattices, with sizes $(m+1)\times m$, $(m+1)\times (m+1)$, $m\times m$ and $m\times(m+1)$. The subsequent contraction follows (b). In this strategy, the largest stored tensor is rank-$\left(\sqrt{N}+1\right)$. The strategies (b) and (c) allow easier parallelism.}\label{fig:Partitioning}
\end{figure*}

Depending on the shape of the lattice, we have developed three different strategies to evaluate the contraction of the tensor netwrok, which are shown in Fig.~\ref{fig:Partitioning}.

In the following we first show a generic way to evaluate the overlap in the main text exactly, namely the equation
\begin{align} \label{eq:overlap}
\langle \vec{\tau} \vert \psi\rangle = \mathcal{F}(\bE_1\bE_2\cdots\bE_N).
\end{align}
Assuming $L_h \leq L_v$,
we first contract all the tensors on the first row to get a rank-$L_h$ tensor
\begin{align}\label{eq:exact1}
F_1^{d_{(1,1)}, d_{(1,2)}, \dots, d_{(1,L_h)}} = \mathcal{F}\left(\left[\bE_{(1,1)}\right]_{d_{(1,1)}}\cdots\left[\bE_{(1,L_h)}\right]_{d_{(1,L_h)}}\right),
\end{align}
where the bottom legs $d_{(1,n)}$ of $\bE_{(1,n)}$ ($1\leq i \leq L_h$) are written explicitly to indicate that they are not contracted in this step. Note also the notation for the position with two numbers instead of one, i.e. $(n,m)$ indicates the qubit on the $n-$th row and $m-$th column.  Next we contract $F_1$ with the first tensor in the second row $\bE_{(2,1)}$ and get
\begin{align}\label{eq:exact2}
G_{1}^{r_{(2,1)}, d_{(2,1)}, d_{(1,2)}, \dots, d_{(1,L_h)}} = \!\!\sum_{d_{(1,1)}} \! F_1^{d_{(1,1)}, \dots, d_{(1,L_h)}}  \times \left[\bE_{(2,1)}\right]_{r_{(2,1)} d_{(1,1)} d_{(2,1)}},
\end{align}
where we have used the fact that for $\bE_{21}$ one has the size $\dim(l_{(2,1)})=1$ and $u_{(2,1)}=d_{(1,1)}$. The resulting tensor $G_{1}$ is a rank-$(L_h+1)$ tensor. Then we contract $G_{1}$ with the second tensor in the second row $\bE_{(2,2)}$ and get
\begin{align}\label{eq:exact3}
G_{2}^{r_{(2,2)}, d_{(2,1)}, d_{(2,2)}, \dots, d_{(1,L_h)}}
=\!\!\!\!\!\!\sum_{r_{(2,1)}, d_{(1,2)}} \!\!\!\!\!\!  G_{1}^{r_{(2,1)}, d_{(2,1)}, d_{(1,2)}, \dots, d_{(1,L_h)}}\left[\bE_{22}\right]_{ r_{(2,1)} r_{(2,2)} d_{(1,2)} d_{(2,2)}},
\end{align}
where we have used the fact that for $\bE_{(2,2)}$ one has $l_{(2,2)}=r_{(2,1)}$ and $u_{(2,2)}=d_{(1,2)}$, and the resulting tensor $G_{2}$ is again a rank-$(L_h+1)$ tensor. We can repeat this procedure and move on to the right until we have contracted all the tensors on the second row and get
\begin{align}\label{eq:exact4}
F_2^{d_{(2,1)}, d_{(2,2)}, \dots, d_{(2,L_h)}} = G_{L_h}^{r_{(2,L_h)}, d_{(2,1)}, d_{(2,2)}, \dots, d_{(2,L_h)}}
\end{align}
where we have used the fact $\dim(r_{(2,L_h)})=1$ and redefined $G_{L_h}$ and $F_2$. Noticing that $F_2$ has the same structure as $F_1$, therefore we repeat this procedure until we have reached the last row and get $F_L$, which is a scalar since all the indexes $\dim(d_{(L_v,n)})=1$ for $1\leq n\leq L_h$. Thus we get
\begin{align}\label{eq:exact5}
\langle \vec{\tau} \vert \psi\rangle = F_{L}.
\end{align}
From this analysis it appears that the largest tensor involved in this procedure is rank-$(L_h+1)$. Moreover, for $L_h>L_v$, instead of moving from top down, it is straightforward to slightly modify the algorithm to move from left to right, and the largest tensor involved would become rank-$(L_v+1)$. Therefore the memory required scales exponentially with the exponent $\min(L_h+1, L_v+1)$.

This generic strategy is shown in Fig.~\ref{fig:Partitioning}(a), where the tensor network is contracted row by row (ideal for a thin lattice where, for instance, $L_v > L_h$). Mathematically, this scheme corresponds to Eqs.(\ref{eq:exact1}-\ref{eq:exact5}). The largest tensor involved in this process is rank-$(L+1)$ where we have defined $L = \min(L_h, L_v)$. Assuming a memory efficient implementation of tensor contraction, one would only require a single tensor of such size since the operand tensor could be overwritten. In the mean time, the most time-consuming step is Eq.(\ref{eq:exact3}), in which one contracts two legs of a rank-$(L+1)$ tensor with two legs of another $4$-dimensional tensor, a process which is repeated $(L_h-2)(L_v-2)$ times. Thus with the contraction scheme in Fig.~\ref{fig:Partitioning}(a), the space and time complexity are
\begin{align}
\spacecpx(L_v\times L_h\times d) &= 2^{\lceil d/8\rceil \left(L+1\right)}, \label{eq:space} \\
\timecpx(L_v\times L_h\times d) &= (L_h-2)(L_v-2)2^{\lceil d/8\rceil (L+3)}. \label{eq:time}
\end{align}
Note that these are very accurate evaluations with a clear prefactor and not just order of magnitude estimates, although the complexities can be reduced by using advanced matrix-matrix multiplications schemes and by parallelizing the operation.

For the special case of a square lattice with $L_h = L_v = \sqrt{N}$, it is possible to improve the performance via a particular partitioning of the sum, as shown in Fig.~\ref{fig:Partitioning}. The partitioning strategies for network with even or odd side length are different, as shown in Fig.~\ref{fig:Partitioning}(b,c) respectively.
For the network with even side lengths, tensors are divided into four parts first, as Fig.~\ref{fig:Partitioning}(b) illustrates. We start the contraction of the tensors from the upper-left partition, obtaining a rank-$\sqrt{N}$ tensor which we refer to as $ F_{ul} $. Similarly, the other three partitions produce another three rank-$\sqrt{N}$ tensors, denoted as $F_{ur}$ (upper-right), $F_{bl}$ (bottom-left) and $F_{br}$ (bottom-right). Then, we contract $F_{ul}$ with $F_{ur}$, and $F_{bl}$ with $F_{br}$. Consequently, by contacting the remaining two tensors together, we get the amplitude value. As a result, the complexity of this strategy is
\begin{align}
\spacecpx(L_v\times L_h\times d) &= 2^{\left(\lceil d/8\rceil\sqrt{N}\right)+1}, \label{eq:spacesquareeven} \\
\timecpx(L_v\times L_h\times d) &= 2^{\left(3\lceil d/8\rceil\sqrt{N}/2\right)+1}. \label{eq:timesquareeven}
\end{align}

The algorithm for the network with odd side lengths ($L_h=L_v=2m+1$) is relatively more complicated. The tensors are partitioned into 4 groups, as shown in Fig.~\ref{fig:Partitioning}(c). The contraction starts from the up-left $(m+1)\times m$ partition, producing a rank-$\sqrt{N}$ tensor denoted as $F_{ul}$. Then we move to the other three parts and contract them into $F_{ur},F_{bl},F_{br}$ same as $F_{ul}$. The contraction of $F_{ur}$ can again be divided into 4 sub-procedures, which are indicated in Fig.~\ref{fig:Partitioning}(c) by the gray dashed lines that break the lattices into 3 small groups. The sub-procedures are: (1) Contracting the right $(m+1)\times m$ tensors into a rank-$\sqrt{N}$ tensor; (2) Contracting the first $m$ tensors at the $m+1$-th column into a rank-$\sqrt{N}$ tensor; (3) Contracting the two rank-$\sqrt{N}$ tensors from procedure (1) and (2) into a rank-$(\sqrt{N}+1)$ tensor; (4) Contracting the obtained rank-$(\sqrt{N}+1)$ tensor with the rank-4 tensor located in the center of the lattice (which is also the left-bottom corner of $F_{ur}$), and resulting in a rank-$(\sqrt{N}+1)$ tensor. Then, by contracting the four parts together, we get the probability amplitude.
As a result, the complexity of this strategy is
\begin{align}
\spacecpx(L_v\times L_h\times d) &= 2^{\lceil d/8\rceil(\sqrt{N}+1)} + 2^{\lceil d/8\rceil\sqrt{N}}, \label{eq:spacesquareodd} \\
\timecpx(L_v\times L_h\times d) &= (2^{\lceil d/8\rceil} +1) 2^{\lceil d/8\rceil(3\sqrt{N}-1)/2}. \label{eq:timesquareodd}
\end{align}

\begin{figure}[htbp]
  \centering
  \subfigure{\includegraphics[width=0.8\textwidth]{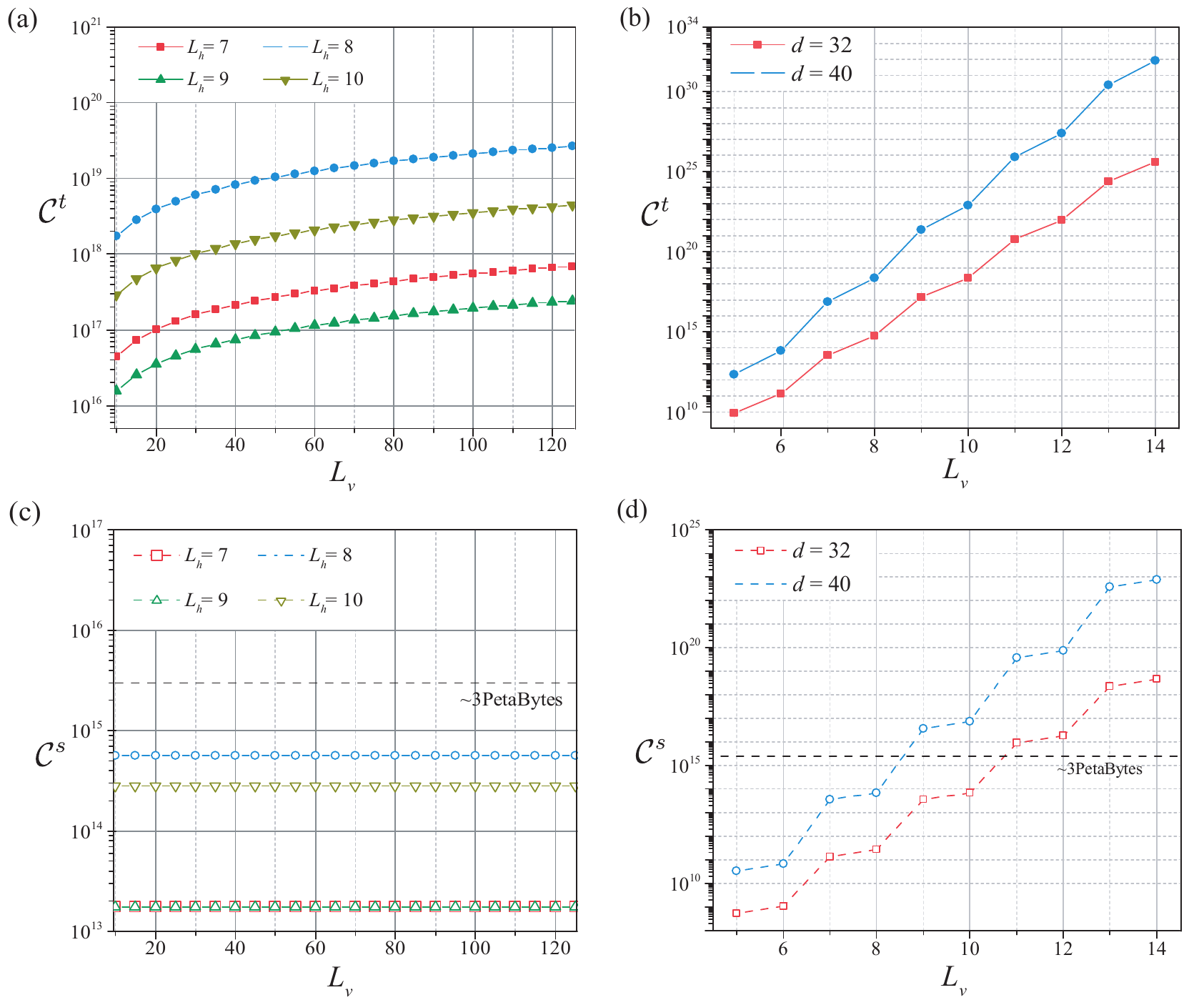}}
  \caption{The space and time complexity of RQCs based on the PEPS quantum circuit simulator for a circuit with $L_h\times L_v$ qubits.
  (a) Scaling of time complexity $\mathcal{C}^t$ with $L_v$ when $L_v > L_h$. (b) Scaling of $\mathcal{C}^t$ with $L_v$ when $L_v = L_h$. (c) Scaling of space complexity $\mathcal{C}^s$ with $L_v$ when $L_v > L_h$. (d) Scaling of $\mathcal{C}^s$ with $L_v$ when $L_v = L_h$. The grey dotted lines in (c) and (d) represent the current memory limit of supercomputers (2.17PB for Tianhe-2 and 2.67 PB for summit).
  } \label{fig:complexity}
\end{figure}

In Fig.~\ref{fig:complexity}(a)
we show the space and time complexities for $8\times l$ circuits for $d=1+40+1$ (or $10\times l$ circuits with depth of $d=1+32+1$), showing that they are within reach for state-of-the-art supercomputers. This shows clearly where the frontier for quantum supremacy stands for this random quantum circuit and for our method. In Fig.~\ref{fig:complexity}(b) we show the space and time complexities computed from Eqs.(\ref{eq:spacesquareeven}-\ref{eq:timesquareodd}).
To this end, we note that our algorithm can be straightforwardly combined with the fast sampling method in~\cite{Villalonga2019,Liu2019} to measure a large number of amplitudes. Following the partitioning strategy, one can sample in one partition with negligible additional cost since the results of the other regions can be reused.

\section{Complexity Analysis of Google Bristlecone QPU}\label{app:Bristlecone}
\begin{figure}
\includegraphics[width=0.35\columnwidth]{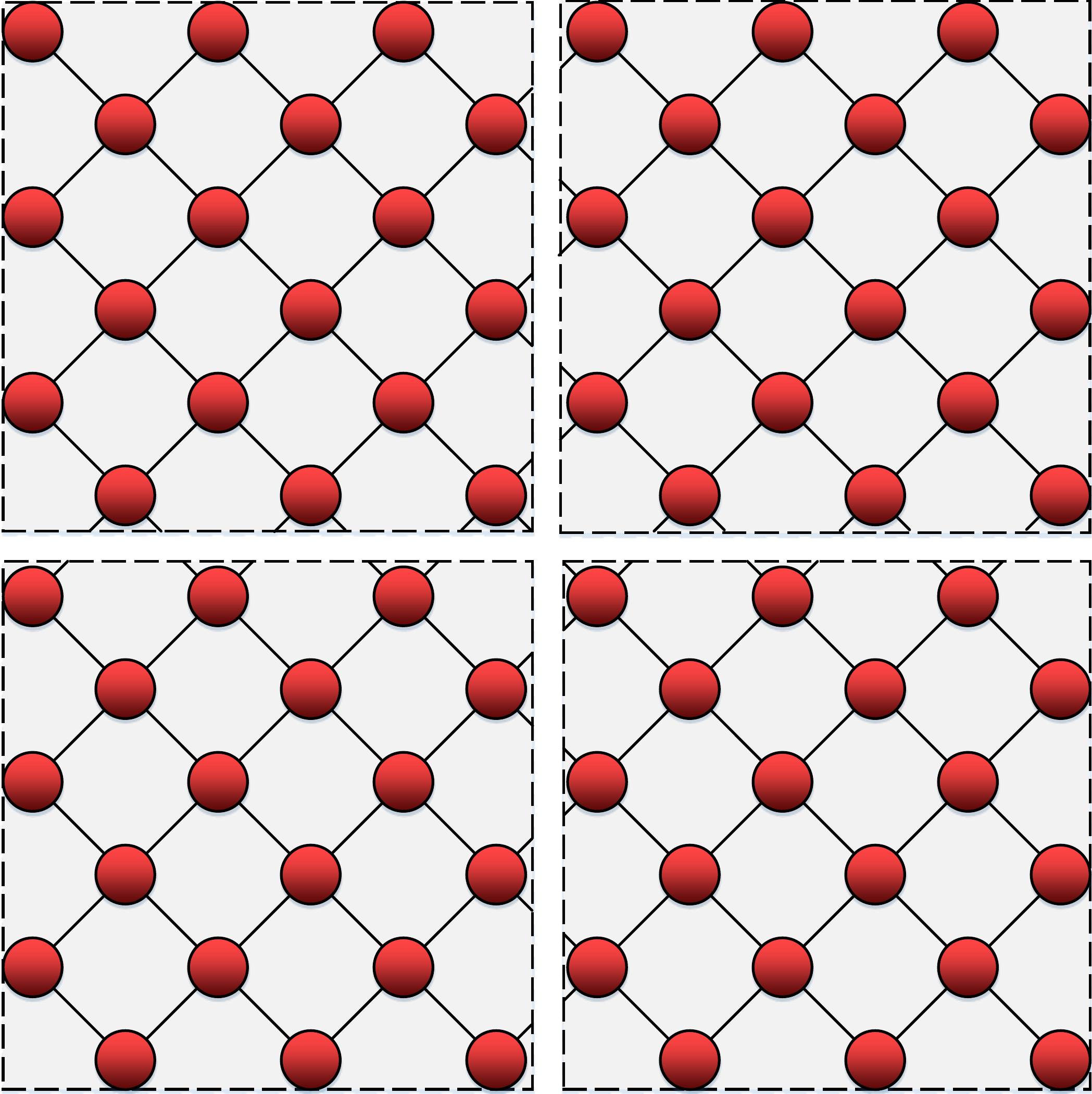}
\caption{Contracting stragety for the Google Bristlecone QPU. The $12\times 6$ lattice is partitioned into four sub-lattices with size $6\times 3$. Contracting all the sub-lattices would result in $4$ large tensors with ranks $11, 10, 10, 11$ (which can be seen by counting the legs which are not contracted.) respectively. Contracting these $4$ large tensors would require to store at least two rank-$11$ tensors.} \label{fig:Bristlecone}
\end{figure}

To simulate the Google Bristlecone QPU with PEPS, both the representation of the quantum state as well as the gate operations are implemeted exactly in the same way as for the rectangular lattice case. The only difference is that during the measurement stage, the tensor network that needs to be contracted are rotated by $45$ degree compared to a rectangular lattice. In Fig.~\ref{fig:Bristlecone} we show a contraction strategy for the simulation of a Google Bristlecone QPU. From Fig.~\ref{fig:Bristlecone} we can see that the number of legs of a tensor is at most $11$, and hence the space cost for simulating this circuit to a depth $(1+32+1)$ with our circuit simulator scales as $2^{32/8\times 11+1} = 2^{45}$, which corresponds to less than $0.6$ PB memory.

\section{Massive Parallel Benchmarking on Supercomputer}

We have implemented our large scale tensor contraction algorithms based on an open-source software package Cyclops Tensor Framework~\cite{Solomonik2014}, with MPI and OpenMP as the parallel interfaces. The massive parallel benchmarking was then executed on Tianhe-2 supercomputer. According to the features of the supercomputer platform and the results of the scaling test, we chose to use one MPI process with 24 OpenMP threads on each node. Each normal node contains two 12-core CPUs, and is equipped with 64GB (128 GB on each fat node) memory. The maximum number of nodes used reaches 4,096 (98,304 compute cores in total), which is less than 1/4 of the whole system, and since we only use CPUs, the peak performance we use is $\sim 1.73$ PFlops. All our calculations are done with double-precision numbers. Our results are listed in Table.1 of the main manuscript.

The numerical simulation with the largest number of qubits is a $10\times 10$ circuit with $d=(1+26+1)$, which is done on 1,024 normal nodes and takes 6 minutes to measure one amplitude, using the partitioning strategy as in Fig.~\ref{fig:Partitioning}(c). The numerical simulation with the largest depth is a $7\times 7$ circuit with $d=(1+40+1)$, which is done on $4,096$ fat nodes and takes 31 minutes. On each fat node 23.13 GB memory is used, and thus this simulation takes 92.51 TB memory in total (detailed data can be found in supplementary information). To pursue efficiency, parts of the data is duplicated on several computing nodes to reduce the cost of data communication, leading to a larger memory usage than theoretical prediction $16$ TB. Here we note that recently in \cite{Villalonga2019} the authors compute, with $0.5\%$ fidelity, $10^6$ amplitudes for a $7\times7\times(1+40+1)$ random quantum circuit with single-precision numbers on Summit in 2.4 hours, using 2.67 PB memory and ${\text R}_{\text {Node-peak}}=200.8$ PFlops. Their optimized implementation, when mapped to unit fidelity, is currently faster than our proof-of-principle calculation.


\bibliography{refs}

\begin{thebibliography}{46}
\bibitem[{iSW()}]{iSWAP}
\bibinfo{note}{If an iSWAP
  $(|00\rangle\langle00|+i|01\rangle\langle10|+i|10\rangle\langle01|+|11\rangle\langle11|)$
  gate is used instead of a CZ gate, then the size of the two tensors after
  factorization will increase by a factor $2$, effectively halving the depth of
  the circuit that can be simulated}.
\end{thebibliography}
\begin{thebibliography}{45}
\bibitem[{iSW()}]{iSWAP}
\bibinfo{note}{If an iSWAP gate $(|00\rangle\langle00|+i|01\rangle\langle10|+i|10\rangle\langle01|+|11\rangle\langle11|)$ is used instead of a CZ gate, then the size of
  the two tensors after factorization will increase by a factor $2$,
  effectively halving the depth of the circuit that can be simulated}.
\end{thebibliography}

\end{document}